\documentclass{aa}
\usepackage{graphicx}
\usepackage{txfonts}
\usepackage{natbib}
\bibpunct{(}{)}{;}{a}{}{,}

\newcommand{\IR}{IRC\,+10\,420~}
\newcommand{\Brg}{Br$\gamma$~}

\def\mathstacksym#1#2#3#4#5{\def#1{\mathrel{\hbox to 0pt{\lower#5\hbox{#3}\hss} \raise #4\hbox{#2}}}}
\mathstacksym\gta{$>$}{$\sim$}{1.5pt}{3.5pt} 
\mathstacksym\lta{$<$}{$\sim$}{1.5pt}{3.5pt} 
\begin{document}
                                %
\title{Resolving the ionized wind of the post-Red Supergiant IRC\,+10\,420 with VLTI/AMBER\thanks{Based on
    observations with VLTI, proposal 077.D-0388}}
\titlerunning{AMBER observations of IRC\,+10\,420}
\author{W.J. de Wit\inst{1}, R.D. Oudmaijer\inst{1}, M.A.T. Groenewegen\inst{2},
                                M.G. Hoare\inst{1}, \and F. Malbet\inst{3}} 
\offprints{W.J. de Wit, \email{w.j.m.dewit@leeds.ac.uk}}
\institute{School of Physics \& Astronomy, University of Leeds, Woodhouse Lane,
                                Leeds LS2 9JT, UK\and
Institute for Astronomy, University of Leuven, Celestijnenlaan 200D, 3001
                                Leuven, Belgium\and
Laboratoire d'Astrophysique, Observatoire de Grenoble, BP 53, 38041 Grenoble,
                                C\'{e}dex 9, France\ 
}          
        
\date{Received date; accepted date}
\abstract
{}
{The paper investigates the milli-arcsecond scale structure
  of the present-day mass-loss of the post-Red Supergiant \IR.} 
{We use three telescopes of the VLT Interferometer in combination with the AMBER
  near-infrared beam combiner to measure spectrally dispersed correlated
  fluxes in the $K$-band around the Br$\gamma$ transition. The resulting
  visibilities are compared to the predicted visibilities of emission
  structures with various simple models in order to infer the size  of
  the observed emission region.}     
{The Br$\gamma$ line is resolved by VLTI+AMBER on all three baselines,
  with the maximum projected baseline extending 69 meter and a
  P.A. ranging between 10\degr~and 30\degr. A differential phase between
  line and continuum is detected on the longest baseline. The Br$\gamma$ emission
  region is found to have a diameter of $3.3$ milli-arcseconds (FWHM),
  when compared to a Gaussian intensity distribution. A uniform disk
  and a ring-like intensity distribution do not fit the line
  visibilities. The size of the continuum emission is not constrained
  by the observations. Comparing the AMBER equivalent width
  of \Brg with measurements from various epochs, we find that the
  stellar photosphere contributes about 60\% of the total
  continuum light at 2.2$\mu$m. The remaining 40\% continuum
  emission is found on scales larger than the 66\,mas AMBER field of
  view. This independently confirms  similar results made by previous
  studies. If the \Brg emission is optically thin, then the observations do not allow to make
any inferences about the shape of the line forming region. However, there
is indirect evidence that the hydrogen recombination line emission is
optically thick. In that case, using simple arguments, we find that the line emitting region
is elongated. This is because the spectrum
indicates that the projected line emitting area is about twice that of the
stellar surface. This value is an order of magnitude less than a circular
\Brg line emitting area would have using the measured size of the emission
region along our baseline. We briefly 
  discuss the possibilities whether such a structure is due to a bi-polar
  flow or a circumstellar disk.}
{}
\maketitle
\section{Introduction}
\label{intro}
Throughout the evolution of a high-mass star, stellar winds carry off a
significant fraction of the initial star mass. This process dominates their
evolution and determines their final fate. In the late stages
of massive star evolution, important information of the mass loss process due
to stellar winds is contained in the complicated circumstellar nebulae, like
the ones found near e.g. LBVs and Wolf-Rayet stars (Nota et al. 1995\nocite{1995ApJ...448..788N};
Nota \& Clampin 1997\nocite{1997ASPC..120..303N}; Van der Sluys \& Lamers
2003\nocite{2003A&A...398..181V}). The nebulae may disclose the mass-loss
history, identifying phases with varying stellar wind properties like the wind
momentum, geometry and total mass lost. One explanation for the observed
nebula geometry is that the different mass loss episodes lead to 
wind-wind interactions, in the sense that a more recent fast wind collides with
a pre-existing slower expanding wind. Crucially, at least one of these winds should be
asymmetric, possibly induced by mass loss anisotropies at the stellar surface
(Maeder 2002\nocite{2002A&A...392..575M}). During which evolutionary phase
this would actually occur is not known however, and various ideas have been
put forward (see e.g. Mellema 1997\nocite{1997A&A...321L..29M}; Heger \& Langer 1998\nocite{1998A&A...334..210H}; Dwarkadas \& Owocki
2002\nocite{2002ApJ...581.1337D}). 

\IR is  one of the few stars that evolve from the Red Supergiant (RSG) phase
back towards the blue, witnessed by an extremely rapid increase in its surface
temperature of $\sim$2200\,K over an interval of 30 years currently having
  a mid-A spectral type (see Oudmaijer et
al. 1996\nocite{1996MNRAS.280.1062O}; Klochkova et
al. 2002\nocite{2002ARep...46..139K}). Its estimated distance of 3.5 to 5\,kpc
implies a luminosity typical for a star with an initial mass of around $\rm 40\,M_{\odot}$, putting
it close to the Humphreys-Davidson limit in the Hertzsprung-Russell diagram (Jones et
al. 1993\nocite{1993ApJ...411..323J}). The star is bright at 2.2$\mu$m
$(K_s=3.6)$ and exhibits a huge infrared excess due to warm dust indicating that
it was only recently in an extreme mass-losing 
Red Supergiant phase (Oudmaijer et al. 1996\nocite{1996MNRAS.280.1062O}). These
properties make the star a prime candidate to eventually become a Wolf-Rayet
star, surrounded by an intricate nebula (Jones et
al. 1993\nocite{1993ApJ...411..323J})   

The circumstellar environment of \IR has various distinctive components. We
list the basic features on the various size scales. Based on HST imaging,
\IR's optical reflection nebula portrays a spherical distribution on scales
larger than 2\arcsec~ out to 6 to 8\arcsec~ (Humphreys et
al. 1997\nocite{1997AJ....114.2778H}). Sphericity of the circumstellar
material on this scale is seen at other wavelengths too, e.g. in the $J$-band by
Kastner \& Weintraub (1995)\nocite{1995ApJ...452..833K}, and in CO transitions
by Castro-Carrizo et al. (2007\nocite{2007A&A...465..457C}). On a scale of
about 1\arcsec, one finds a bipolar geometry, with two patches of $K$-band and
mid-IR emission to the NE and SW direction of the star (Humphreys et al. 1997). Given the relatively
large distance from the central star, this emission should be scattered
radiation. The location of these patches corresponds to an approximate
position angle of 40\degr. Closer in and down to 0.3\arcsec, HST reveals a
complicated and non-uniform distribution of various jet-like, ray and arc
features (Humphreys et al. 1997).

\begin{table*}
  {
    \begin{center}
      \caption[]{Predicted and measured continuum visibilities for \IR and the
	calibrator star HR\,7648. \IR has an observed upper limit to its
	diameter of 3\,mas (Monnier et al. 2004), whereas the calibrator has a
	uniform disk diameter of 1.95\,mas (Richichi et al. 2005). Corresponding
	theoretical visibilities are given in cols.(2) and (6). The measured raw
	values of $\rm V_{cont}^{2}$ in cols. (3) and (7) correspond to the
	average continuum value for a SNR frame selection of 20\% (see text),
	these values are subject to systematic uncertainties. The uncertainties
	on the raw $\rm V_{cont}^{2}$ given reflect the internal
	variation. Col.(8) is the calibrated $\rm V_{cont}^{2}$. Col.\,(9) gives
	the ratio between the line and continuum (see
	Fig.\,\ref{select}). Col.\,(10) is the range for $\rm V_{Br\gamma}$ as
	derived in Sect.\,\ref{brgsize}, taking into account the various
	uncertainty contributions.}
      \begin{tabular}{cccc|ccccccc}
        \hline
        \hline
        &  HR7648     &              &                    & \IR      &       &         &                     &                      &                  & \\
        &  Baseline   & $\rm V_{UD}$ & $\rm <V_{cont}^2>$ & Baseline & P.A.  &  V      & $\rm <V_{cont}^2>$   &   $\rm <V_{cont}^2>$  & V$^2_{\rm cont.}$/V$^2_{Br\gamma}$  &   $V_{Br\gamma}$ \\
        &  (m)        & theo.         & raw                & (m)      & \degr &  theo.  & raw  & calib.   &&\\
	&  (1)  & (2) & (3) & (4) & (5) & (6) & (7) & (8) & (9)  & (10) \\
        \hline        
        U1-U2 &  43.9 &  0.96  & $0.111\pm0.002$          & 39.8     & 12.8  & $>0.92$ & $0.169\pm0.004$ & $1.46\pm0.03$& $1.36\pm0.07$&0.60--0.74\\
        U2-U3 &  35.8 &  0.97  & $0.149\pm0.003$          & 30.3     & 32.7  & $>0.95$ & $0.210\pm0.005$ & $1.37\pm0.03$& $1.20\pm0.07$&0.74--0.89 \\
        U3-U1 &  78.7 &  0.87  & $0.201\pm0.004$          & 69.1     & 21.5  & $>0.76$ & $0.265\pm0.005$ & $1.15\pm0.03$& $1.68\pm0.06$&0.39--0.57\\
        \hline 
      \end{tabular}
      \label{tabvs}
    \end{center}
  }
\label{obs}
\end{table*}

At even smaller angular scales, the dusty environment giving rise to the near-IR
excess has been resolved with speckle interferometry at 2.11$\mu$m (Bl\"{o}cker
et al. 1999\nocite{1999A&A...348..805B}). Modelling the spectral energy
distribution (SED) and visibility curves with DUSTY leads these authors to
conclude that \IR is surrounded by two separate spherical shells with diameters
of $0.070\arcsec$ and $0.310\arcsec$ centred on the star. Finally, the
geometry on a size scale comparable to the stellar radius ($\sim$
milli-arcseconds) can be probed with the various emission lines, like H$\alpha$
and \ion{Fe}{ii}, that are reflected off the large scale blue nebula (Humphreys
et al. 2002\nocite{2002AJ....124.1026H}; Davies et al. 2007\nocite{2007arXiv0708.2204D}). With IFU
observations of the nebula, Davies et al. find evidence for a present-day wind
which has an axi-symmetric, rather than spherically symmetric geometry. The
H$\alpha$ emission symmetry axis is found to have a position angle of 33\degr.

In this paper on \IR we present complementary observations of the subarcsecond
ionised and continuum emitting material inside the 70\,mas dust shell found by
Bl\"{o}cker et al. (1999\nocite{1999A&A...348..805B}). These authors found that 
about 60\% of the $K$-band continuum emission remains unresolved (see also
Monnier et al. 2004), presumably corresponding to the star itself. We use the VLT 
interferometer in conjunction with the near-IR three telescope beam combiner
AMBER, attaining a maximum spatial resolution of 6.5\,mas. The ultimate objective is
to measure the size and shape of the Br$\gamma$~emission line region, giving a
direct measurement of the present-day mass-loss geometry, an important
ingredient to the mass-loss puzzle. In Sect.\,\ref{observ} we present the
AMBER data, and provide a list of steps and considerations taken in
the reduction process. Results regarding the final visibility, flux and
differential phase spectra of \IR are presented in Sect.\,\ref{results}. In
this section we derive  
the size for the continuum and Br$\gamma$~emission. We discuss the shape
of the Br$\gamma$~emission region in Sect.\,\ref{discu}. The latter is found to be
elongated along the VLTI baseline of the observation with a P.A. of $\sim 20\degr$. We
conclude our findings in Sect.\,\ref{concl}.

\section{Observations and data reduction}
\label{observ}
\subsection{AMBER data reduction: general considerations}

AMBER is the VLTI's near infrared, three telescope beam combiner (Petrov et
al. 2007\nocite{2007A&A...464....1P}). The interferometric beam is fed into a spectrograph
delivering a spectrally dispersed interferogram that is spatially encoded on a
detector. AMBER can operate with two or three telescopes, and offers three
spectral resolutions. The AMBER data reduction is not (yet) a straightforward
process and particular care should be taken in reducing and analysing the
data.

\begin{figure}
  \includegraphics[height=8cm,width=8cm]{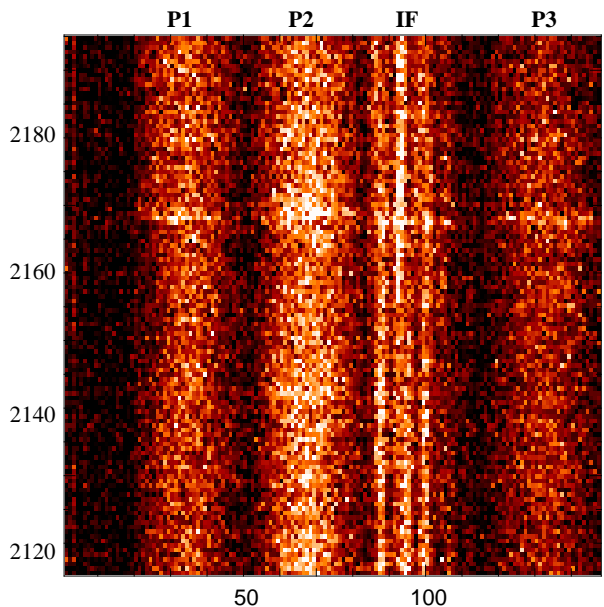}
  \caption[]{Illustration of an AMBER frame of the target star
  IR\,+10\,420 showing fringes. An approximate wavelength scale in nm
  is given along the y-axis, pixels along x-axis. The frame gives the
  three telescope channels (P1, P2, P3), and the interferometric
  channel (IF). The Br$\gamma$ emission line can be identified as the
  horizontal enhancement in recorded ADUs.}
  \label{frame}
\end{figure}

The reduction of AMBER data can be split in two distinctive steps. Ever since
the discovery of an electronic interference pattern on the detector (Li Causi et
al. 2008), a first reduction step must involve a
digital filtering process that needs to be applied to all data taken with
AMBER. A software routine by the name of AMDC has been developed that identifies
the electronic interference frequencies in Fourier space by means of a dark
frame, and creates a ``Fourier mask''. This mask needs to be applied to the
data, and AMDC replaces the power at the corrupted frequencies with a new value
based on the power at the neighbouring frequencies.

The second step consists of converting the Fourier filtered AMBER data to
(raw) visibilities.  This step can be done using the dedicated AMBER software
{\it amdlib} written by the consortium (see Tatulli et al. 2007\nocite{2007A&A...464...29T}
for an in depth explanation). In short, the software algorithm relies on a
mathematical description of the interferogram, with empirical input from the
calibration and alignment unit of the AMBER instrument. It records
VLTI$+$AMBER instrumental characteristics at the beginning of each night, and
they are used to create the so-called pixel-to-visibility matrix (P2VM). The
P2VM converts subsequently the observed science target fringe patterns into a
measurement of coherent flux. Alongside the contemporaneous registration of
the total photometric flux spectrum at each telescope for each
interferogram recorded, the data reduction package calculates the
interferometry observables, i.e. in case of three telescopes, the three
squared visibilities, closure phase, and differential phase.  An example of an
AMBER frame of \IR is given in Fig.\,\ref{frame}. It consists of three
telescope channels and an interferometric channel. In this figure an
approximate wavelength scale is given on the vertical axis (no wavelength
calibration has been applied). The presence of Br$\gamma$ emission is
identified by the bright horizontal line, crossing the three photometric and
interferometric channels.

\subsection{Observations with AMBER}

VLTI+AMBER observed \IR on the nights of 14th and 15th June 2006. The
observations were performed in the UT1-UT2-UT3 baseline configuration. UT4 was
not available during these nights due to technical problems. The telescope
configuration delivers projected baselines between $\sim 40$ and $\sim 70$
meters and a limited P.A. coverage of about 20\degr~(see Table\,\ref{tabvs}). The
AMBER instrument was set-up in the medium spectral resolution mode,
delivering a spectral resolution of 1500. This chosen wavelength range lies
between 2.12 and 2.19$\mu$m, and includes the Br$\gamma$
transition. Br$\gamma$ has been reported to be in emission by Oudmaijer et
al. (1994\nocite{1994A&A...281L..33O}) and the size of the emitting region is the goal of our
AMBER observations of \IR. 

On the first night a total of 15 observations of \IR was secured over a period
of one hour under reasonable weather conditions. On the second night 5
observations were secured under bad weather conditions.  Each observation with
AMBER produces a file that consists of 1000 individual exposures, called
frames. Each frame is then made up of the individual photometric information
for each telescope aperture and the interferometric beam information (the
interferogram), all as function of wavelength (see Fig.\,\ref{frame}). The
integration time per frame was chosen to be 50~ms. The integration time is
limited by the movement of the fringes over the detector by atmospheric
turbulence (the so-called jitter), causing attenuation of the fringe
visibility.

In order to perform an absolute calibration of the AMBER visibilities, an
interferometric calibrator star was observed 5 times each night. The
calibrator star is catalogued as \object{HR\,7648} (SAO\,125355). It is a
$K=2.2$, K5\,III star with a uniform disk diameter of $1.95\pm0.02$ mas
derived from spectrophotometric data (Richichi et
al. 2005\nocite{2005A&A...431..773R}). The ASPRO tool\footnote{ASPRO is a java
applet for the preparation and simulation of VLTI observations; see
http://www.mariotti.fr/aspro\_page.htm.} returns a size of $2.056\pm0.142$mas
for HR\,7648 based on magnitude-colour calibration (like $V$ vs $V-K$) of
angular diameters. The calibrator's size is thus not directly measured using
interferometry. 
The object was chosen as it is close on the sky to \IR and it is expected to
remain (nearly) unresolved (see Table\,\ref{tabvs}). Note, that the calibrator
is considerably brighter than the expected correlated magnitude of the science
target (60\% of $K=3.6$, Monnier et al. 2004\nocite{2004ApJ...605..436M}). The
calibrator was observed 5 consecutive times, one hour after the first \IR
observation, with the same exposure time as the target. The
observation of the calibration star was done under worse seeing ($1.1\arcsec -
1.2\arcsec$) conditions than the target star ($0.8\arcsec$), which implies
better AO correction during the target observations.

All AMBER data were corrected for electronic interference noise using the AMDC
software by Li Causi et al. (2007), and reduced using {\it amdlib} (versions
1.21 and 2.0 beta) software with frame bin size 1, i.e. no averaging over the 1000 frames
present in each file. The
additional data reduction files, namely a bad pixel map and a flatfield were
taken from the AMBER website (versions of 9 Feb 2006) and used in the reduction
process. Visual inspection using the ammyorick environment (see e.g. Millour et
al. 2007\nocite{2007arXiv0705.1636M}) shows that only the first five \IR
observations do actually contain frames with fringes. 

A measure of successfully interfering the three telescope beams and recording
of fringes is given by the amdlib package parameter signal-to-noise ratio
(SNR).  The expression for this parameter can be found in eq.~(20) of Tatulli et
al. (2007a\nocite{2007A&A...464...29T}). It is based on the weighted summing of
the correlated flux over all spectral channels per interferogram. 
There is a strong need to apply a data quality (fringe quality) check to AMBER
data due to non-stationary vibrations in the interferometer infrastructure,
which for the moment are uncontrollable. The currently recommended approach
regarding these data quality degrading effects is by making a frame selection
based on the SNR criterion, and subsequently disregarding low SNR frames up to
a certain cut-off (e.g see Tatulli et al. 2007b\nocite{2007A&A...464...55T}). 
The SNR values calculated by {\it amdlib} for these observations range between 1
and 4. The other ten observations of the first night suffer from very low
fluxes, hardly rising above the detector noise level. The five calibrator
observations are all of relatively good quality, with some frame SNR values
reaching 10. Observations of the second night are also of low quality. The
results presented in this paper are therefore based on the first five
observations of both \IR and HR\,7648 taken during the first night.

\begin{figure*}
 \center{\includegraphics[height=14cm,width=14cm]{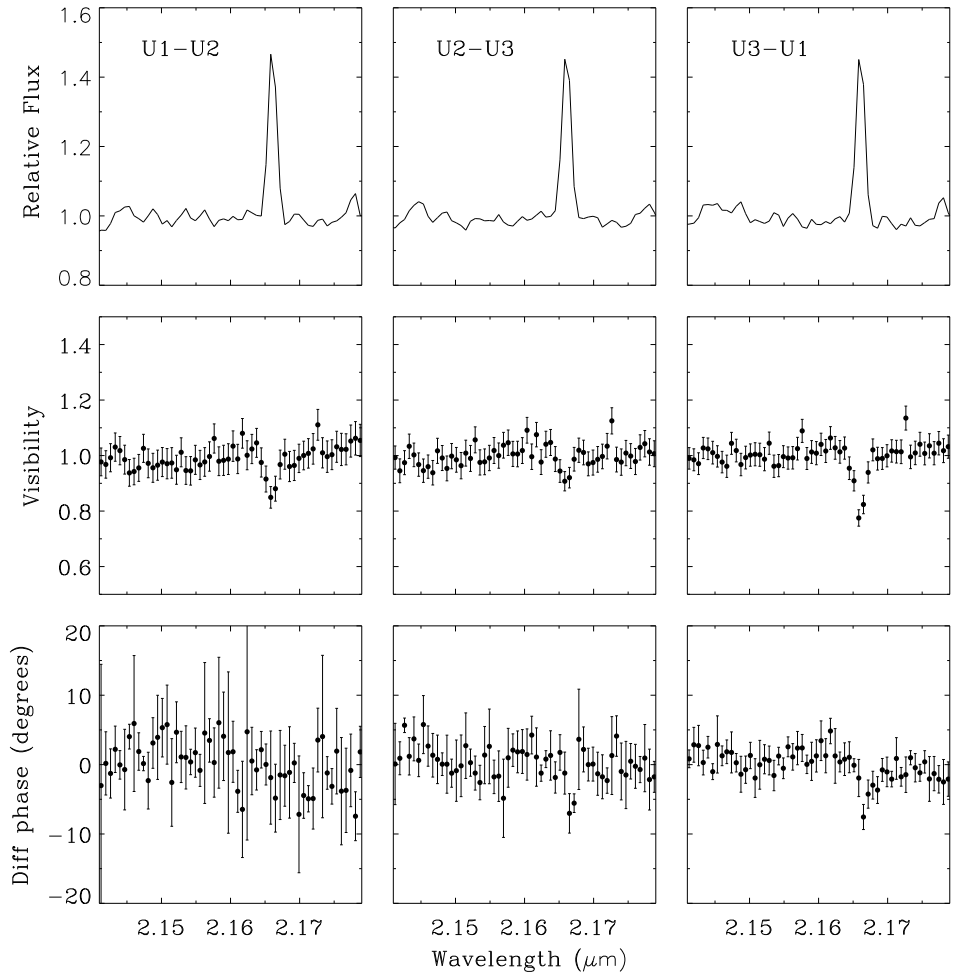}}
  \caption[]{AMBER flux (top), visibility (middle) and differential phase
    (bottom) spectra of \IR for each of the three VLTI U1-U2-U3
    configurations. The flux spectrum is divided by a telluric standard and normalized to
    one. The continuum visibility is normalised to a value of one near \Brg (see
    discussion in text). The Br$\gamma$ emission line has a smaller visibility
    than the continuum at all three baselines. A phase change is
    detected for the U3-U1 baseline.} 
  \label{VB}
\end{figure*}

Practical experience has led to the compromise to choosing the best 20\% of the
frames during the frame selection process (Tatulli et al. 2007a). This 20\% SNR
cut-off was chosen only by the consistent AMBER visibility results it has
provided (see e.g. Malbet et al. 2007\nocite{2007A&A...464...43M}). In our case
the 20\% selection for calibrator and target would lead to unphysical calibrated
target continuum visibilities larger than one ($<V^{2}_{\rm cal}>$ in
Table\,\ref{tabvs}). However, we shall show in the next subsection that frame
selection does not seriously affect the {\it relative} AMBER visibilities.

The current quality of AMBER data limits the ability to perform an absolute
calibration of the continuum visibilities. Instead, we will use the
results from previous high-angular resolution studies in the literature. We
quote in Table\,\ref{tabvs} an upper limit to the diameter of \IR's continuum
emitting region of 3\,mas based on Keck Interferometer observations in the
$K$-band (Monnier et al. 2004). We have listed the corresponding visibility
lower limits for a uniform disk for each VLTI baseline. We note that 3\,mas is
probably a generous size upper limit for the continuuum emitting region, if
it corresponds to the underlying star only. A simple calculation shows that for
a luminosity of \IR of $\rm 25\,462~(d/kpc)^2~L_{\odot}$ (Bl\"{o}cker et
al. 1999) and a mid A spectral type (Oudmaijer et al. 1996), a much smaller
stellar diameter of $\sim 0.70$\,mas can be expected. This result is
independent of distance. If instead the stellar disk of \IR were equal to the
$\sim 3$\,mas upper limit, then its spectral type would consequently need to be
M-type, which is inconsistent with \IR's A-type spectrum.  As we shall argue
below, other contributions to the continuum as seen by AMBER by dust or any
bound-free/free-free emission is small.

\section{Results}
\label{results}
\subsection{AMBER flux, visibility, and phase spectra}
\label{vispec} 
The panels of Fig.\,\ref{VB} show the extracted AMBER observables for each UT
baseline: the flux (top), visibility (middle) and the differential phase
(bottom) spectra.  The flux spectra are normalised to one (at Br$\gamma$). No
explicit wavelength calibration has been performed, and the 
spectra have simply been shifted in wavelength in order to match the emission
line with the Br$\gamma$ rest wavelength. The spectra have been divided by the
interferometric calibrator star in order to remove telluric absorption
components. The flux spectra show the Br$\gamma$ transition clearly in emission,
as has been reported by three other papers previously (Oudmaijer et
al. 1994\nocite{1994A&A...281L..33O}; Hanson et
al. 1996\nocite{1996ApJS..107..281H}; Humphreys et al. 2002\nocite{}), with
spectra taken in 1992, 1994, and 2000 respectively. The line has an equivalent
width (EW) of $-7\AA$, which is significantly stronger than found in the above
mentioned studies. We discuss this difference in the next subsection.

The second row of three panels present the visibility spectra. The spectra are
the average of 20\% best SNR frames. Following the discussion on the
calibration of the continuum visibilities, we have set its value to one,
i.e. a spatially unresolved continuum (see also next subsection). The important result of these observations is
however the smaller visibility in the Br$\gamma$ emission line as function
of baseline. We note that the shortest baseline is the U2-U3 baseline, where
the deviation from a spatially unresolved \Brg line is smallest.

\begin{figure}
  \includegraphics[height=8cm,width=8cm]{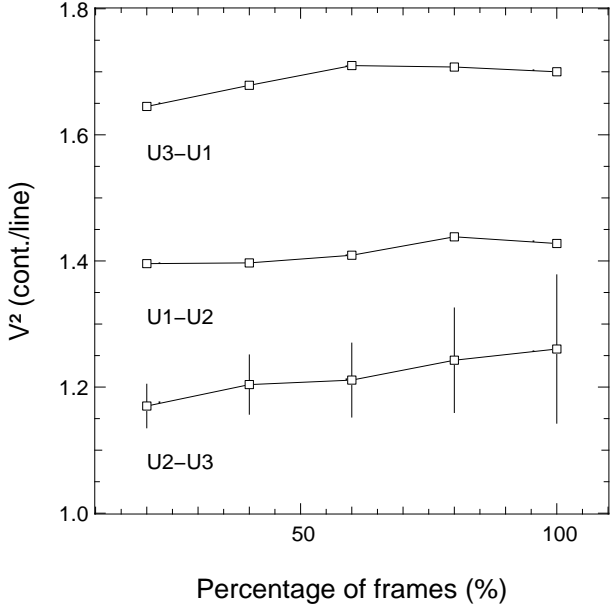}
  \caption[]{The ratio of the continuum squared visibility to the Br$\gamma$
  squared visibility as function of the number of frames selected.The errorbars 
  indicate the continuum variance per spectrum averaged over the selected
  frames. For clarity purposes, the errorbars are plotted for one baseline only.} 
  \label{select}
\end{figure}

Crucial for the relative  visibility at the Br$\gamma$ line with respect to
the continuum is its stability as function of the frame selection criteria. 
To test this, we apply a SNR frame selection with an increasing percentage of frames
selected and for each selection we calculate the relative visibility of the
Br$\gamma$ line. We start with a minimum percentage of 20\% and increase it with
steps of 20\%. For each selection we measure the average of the continuum (excluding
the Br$\gamma$ transition). The minimum in the Br$\gamma$ transition is taken to
be the minimum of a Gaussian profile fit to the resulting visibility spectrum. The evolution
of the fraction $V^2_{\rm cont}/V^2_{\rm Br\gamma}$ as function of percentage of
selected frames is shown in Fig.\,\ref{select}. The figure shows that this
fraction slightly increases with the percentage of frames selected. A bias
towards smaller  values of $\rm V^2$ is introduced when selecting more frames
explains the observed trend. This is because a correct estimate of $\rm V^2$ requires a certain level
  for the incoming flux. With Br$\gamma$ in emission, more line flux is
  available for beam combination than in the adjacent continuum. A reasonable estimate for $V^2_{\rm
  Br\gamma}$ exists therefore for a larger fraction of the frames than for the $V^2_{\rm
  cont}$ estimate. Increasing the number of selected frames will thus bias the
$V^2_{\rm Br\gamma}$ less than it does $V^2_{\rm cont}$, resulting in a larger value for
their ratio. The errorbars on the measurements in Fig.\,\ref{VB} are
representative of the variance of the continuum visibility level, averaged over the
indicated number of selected frames. The figure shows 
that the strongest relative dependence on frame selection is found for the shortest baseline 2
(U2-U3), where Br$\gamma$ has the smallest deviation from the continuum. The
relative visibility changes at most with 0.1 for this baseline. The variation of these fractions are taken
into account in the uncertainty in the ratio between line and continuum as 
quoted in Table\,\ref{tabvs}. We thus conclude that the AMBER relative visibilities
of \IR are stable within the uncertainties listed. 

Finally, the last row of panels in Fig.\,\ref{VB} shows the differential
  phase for each baseline. The differential phase is an approximate measure for the
  spatial offset of the photocentre on the projected baseline at a certain
  wavelength relative to some reference wavelength.  The data show no change in
  differential phase over the Br$\gamma$ line on the first two baselines (
  but perhaps a trace is present on the second baseline). A phase
  change of 5 to 10 degrees in the Br$\gamma$ line with respect to the continuum is detected on the
  third baseline. This constitutes an offset of the Br$\gamma$ photocentre of
  about 0.1\,mas, and will be discussed in Sect.\,\ref{brgsize}.
 
The fourth AMBER observable, the closure phase, shows a random
white noise scatter with an rms of 25\degr \, around 0\degr \, (for 20\% best
SNR frame selection), without any trace of a signal around the Br$\gamma$
transition.

\subsection{Size scales of the {\it dusty} CS environment}
The total $K$-band emission of \IR appears to consist of almost 
 equal contributions by the photosphere and the hot CS dust (Oudmaijer et
al. 1996). They find a good SED fit for a star surrounded by two
  dust shells at distances larger than 86 stellar radii (30\,mas).  The
  Bl\"{o}cker et al. (1999) interferometric observations resolve $\sim40\%$ of
  the $K$-band emission, and subsequent modelling of both the SED and
  visibilities reveals two dust shells with a minimum radius $\sim35$\,mas (see
  also Sect.\,\ref{intro}).  The 60\% flux component remains unresolved even for
  baselines up to 36 meters as probed with IOTA (Monnier et
  al. 2004\nocite{2004ApJ...605..436M}). Evidently, most of the $K$-band excess corresponds
  to this dust component, leaving at most 10\% of the {\it total}
  continuum due to another source. This could be either dust or gas emission located much
  closer to the star and unresolved by IOTA. The relatively large $\rm
  EW(Br\gamma)=-7\AA$ found from the AMBER flux spectra is consistent with the idea
  of a 40\% reduction of the $K$-band continuum due to a field of view effect rather than variability
  in the amount of \Brg flux.  Previous observations reported $\rm EW(Br\gamma)$
  of $-1.2\AA$ (Oudmaijer et al. 1994) and $-4\AA$ (Humphreys et al.  2002).
The question is how much the \Brg emission has varied between the 2000 spectrum
by Humphreys et al. and the AMBER spectrum of 2006. An as yet unpublished
spectrum taken by our group at the NTT with the SOFI imaging spectrograph in
2005, has an EW(Br$\gamma$) of $-4.5\AA$, i.e. hardly any change with respect to
2000. This is in line with EW(H$\alpha$) that varied relatively little over that
same period (Patel et al. 2008).  Assuming thus no change in \Brg emission, we conclude
that $\sim$40\% less continuum must have been observed in the AMBER data, which 
indicates that most dust is outside AMBER's field of view (66\,mas).

\subsection{Size scale of the {\it Br$\gamma$} emission region}
\label{brgsize}
The size of the Br$\gamma$ emitting region can be derived from the line
visibilities. We estimate the visibility of the Br$\gamma$ emission component
by applying the definition for the total visibility of a multi-component
source :
\begin{equation} V_{\rm Br\gamma+cont} = \frac{V_{\rm cont} \times F_{\rm
cont}+V_{\rm Br\gamma} \times F_{\rm Br\gamma}}{F_{\rm cont}+F_{\rm
Br\gamma}}.
\end{equation} 
We solve for $V_{\rm Br\gamma}$ using the observed line
and continuum flux ratios from the spectrum. However in this procedure there
are uncertainties involved regarding the absolute level of $V_{\rm cont}$ as
previously discussed and the value of $F_{\rm cont}$. $F_{\rm cont}$ is
uncertain in the sense that at Br$\gamma$ the proper underlying photospheric
absorption profile has to be used. We recall that the ratio of $V_{\rm cont}$
to $V_{\rm Br\gamma+cont}$ is stable as shown in Sect.\,\ref{vispec}, $V_{\rm
cont}$ however could in principle lie between the values corresponding to the
3mas upper limit from IOTA, or be fully unresolved. 

Eq.\,(1) is valid if there does not exist a differential phase between
line and continuum (see e.g. Weigelt et
al. 2007\nocite{2007A&A...464...87W}). The
effect of the detected U3-U1 differential phase on the line visibility is
marginal compared to other uncertainties that we will discuss next.

\begin{figure}
  \includegraphics[height=8cm,width=8cm]{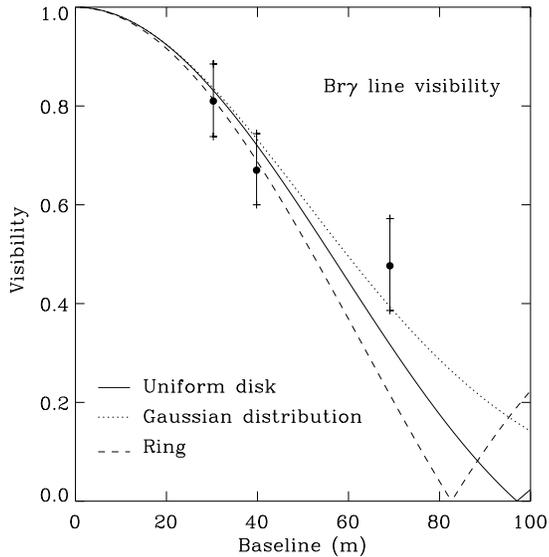}
  \caption[]{The final visibilities of the \Brg emitting region. The range in
  values is determined by the uncertainty in photospheric absorption, the
  absolute continuum visibility, and the error on the relative visibility
  of the line. Three simple, symmetric models have been fit to the data. The
  Gaussian model fits with a FWHM of 3.3\,mas.}
  \label{Brg}
\end{figure}

First, we estimate the value for $F_{\rm cont}$ in eq.\,(1) that needs
to correspond to the photospheric absorption feature. We estimate the
absorption feature from the spectral type of \IR, {\it viz.} mid
A-type. We introduce into the AMBER spectrum a Gaussian absorption
profile with EWs of 5 and 6\AA~with a FWHM similar to the emission
line. This EW is about the average for A-type Ia supergiants as
reported in Hanson et al. (1996\nocite{1996ApJS..107..281H}). The
ratio of $F_{\rm Br\gamma}$ to the adjacent continuum is 1.48, whereas
the ratio of $F_{\rm Br\gamma}$ to the minimum flux level of the
photospheric absorption profile is between 1.97 and 2.08 for an
equivalent width of 5\AA~and 6\AA~respectively. Given the discussed
uncertainties in the continuum visibilities, we derive a range of
values for the Br$\gamma$ visibility at each baseline. The derived
values are presented in Fig.\,\ref{Brg} and listed in Table\,\ref{obs}.

In the figure we fit the Br$\gamma$ visibilities to three simple
geometrical light distributions, {\it viz.} the uniform disk, the
Gaussian distribution, and a ring distribution. The uncertainty is
dominated by three factors, (1) uncertainty of the absolute
calibration of the continuum visibility, between 3\,mas and
unresolved, (2) uncertainty of the relative strength of \Brg
visibilities relative to the continuum (see Table\,\ref{tabvs}
col.\,(9)), and (3) uncertainty regarding the underlying photospheric
absorption. All these three uncertainties are taken into account in
the range of \Brg visibilities plotted in Fig.\,\ref{Brg} and listed
in col.\,(10) in Table\,\ref{tabvs}.  The figure clearly demonstrates
that the shapes of the visibility curves as function of baseline do not
follow the observed visibilities in case of a uniform disk and a
ring. On the other hand the Gaussian distribution can reproduce the
observations, albeit marginally. The Gaussian distribution of
Fig.\,\ref{Brg} has a FWHM of 3.3\,mas. We thus conclude that the
Br$\gamma$ emitting region on the sky of \IR takes the shape of a
Gaussian-like distribution with a FWHM of $\sim 12$\,AU ($d=3.5$kpc),
at least between the measured baseline position angles, i.e. between a
P.A. of $\sim 10\degr$ and $\sim 30\degr$.

\section{Discussion: Geometry of  the {\it Br$\gamma$} emission region}
\label{discu}
The shape of the emitting region can not be further constrained by the
visibility measurements alone, given the limited P.A. range of our
VLTI observations. Only when the line is optically thick can we infer
 a geometry by means of the total \Brg flux and the total continuum flux
for an equivalent wavelength range (taken to be the 
full width of the emission line). Indeed, in that case, the flux will be
  approximated by $F_L = B_{\nu}(\nu,T) \Delta \nu \left( 
\frac{R}{d} \right)^2 $ (cf. Oudmaijer et al. 1994; see also
Grundstrom \& Gies 2006\nocite{2006ApJ...651L..53G}), with $B_{\nu}$
the Planck function, $\Delta \nu$ the linewidth, unresolved in our
data but measured to have a FWHM of 64 kms$^{-1}$ by Oudmaijer et
al. (1994) and only slightly broader in Humphreys et al. (2002), $R$
represents the radius of the emitting surface if it is circular, and $d$ is
the distance. The line optical depth has no bearing on the size
of the \Brg emitting region as  determined in the previous section.

Oudmaijer et al. (1994) present the dereddened linefluxes of several
optical and near-infrared hydrogen recombination lines and demonstrate
that the line ratios deviate significantly from case B. For example,
\IR shows $F_{Br\gamma} \sim F_{Br \alpha}$, a line ratio that
indicates that Br$\alpha$ is optically thick (case B predicts a ratio
of 0.33). This would not necessarily imply that Br$\gamma$ is
optically thick of course, but we note that the models by Simon et
al. (1983\nocite{1983ApJ...266..623S}) indicate that Br$\gamma$ is
almost as strong as Br$\alpha$ for a very large range of optical
depths for both Br$\gamma$ and Br$\alpha$. As both H$\alpha$ and the
Br$\gamma$ line have become stronger over the years (see Patel et
al. 2007 and above respectively), we conclude that it is reasonable to
assume the line to be optically thick. 

Inspection of the above equation immediately indicates that we can
directly compare the size of the emitting region with that 
of the star, emitting at the same wavelength by taking the ratio

\begin{equation}
\left( \frac{F_L}{F_*} \right) = \left( \frac{B_{\nu}(\nu,T_L)}{B_{\nu}(\nu,T_*)} \frac{A_L}{A_*} \right)
\end{equation}
With $T_L$, $T_*$, $A_L$, $A_*$, the temperature and the projected
surface area of the line emitting region and the star respectively.
In the AMBER data, taking into account filling in of the photospheric
absorption profile by the line emission, we find that the emission
lineflux $F_L \sim F_*$.  Assuming a stellar temperature of 8500 K and
a temperature of the ionized region of 5000 K, we can then use Eq. 2
to compute the ratio of line emitting surface area and the stellar
disk, which turns out to be 1.7.  If we take the same temperature for
the star as for the line emitting region, we find a lower limit to the
ratio of 1. So, the flux spectrum indicates that an optically thick line emitting
area would be less than twice the size of the stellar photosphere. 

How does this compare with the results from the AMBER interferometry?
We recall that the Gaussian fit to the visibility of the resolved
Br$\gamma$ line gave a FWHM for the Br$\gamma$ line emitting area at
our projected baseline of 3.3 mas. This diameter is 5 times larger
than the stellar diameter as derived from spectral energy distribution
considerations and even larger than the upper limit of 3 mas found by
Monnier et al. Hence, if the optically thick Br$\gamma$ came from a circular region on
the sky, then based on the observed and inferred size scales of the
line emitting region and the star respectively, we would predict the
line flux to be at least an order of magnitude larger than is actually
observed.  This apparent contradiction between observed and expected
line fluxes can  most straightforwardly be explained if the line
emitting surface is not circular, but has an asymmetric, elongated,
appearance, with the measured size scale along the major axis. 

Such an apparent disparity is not unusual, and is a well documented 
property of Be stars. Their, optically thick, H$\alpha$ lines are
known to originate from a disk and the line fluxes are also comparable
to the continuum flux, implying small emitting areas. However, in the
case where the emission is resolved, the disks' major axes are much
larger than the unresolved stars. Indeed, the effect is so general
that the H$\alpha$ EW can be used to estimate the extension of the
disks (see e.g. Grundstrom \& Gies 2006).

In summary, the \Brg line visibilities indicate that AMBER resolves the line
emitting region, and we derive a corresponding size of $\rm 5\,R_{*}$.  The region
produces a much larger line-to-continuum flux ratio than observed in the AMBER
flux spectrum, under the assumption of optically thick emission.
From this discrepancy one may infer that the \Brg
emitting region is actually elongated on the sky along the North-Eastern
direction, with a P.A. of about 20\degr. A caveat is of
  course that for an optically thin \Brg line no such conclusions can be made
  regarding the  geometry of the emitting region.

That the circumstellar material of \IR exhibits a North-Eastern extension at
  larger scales than probed here may not be a coincidence. The outer 
reflection nebulosity in the HST 
images of Humphreys et al. (1997) is elongated along a PA of 33\degr , while
Davies, Oudmaijer \& Sahu (2007) find from a study of the reflected optical
spectrum off the dust, that the H$\alpha$ emission has an axis of symmetry
aligned with the extended reflection nebula. Although their data do not allow any
conclusions to be drawn about whether the H$\alpha$ emission is elongated along
or perpendicular to the long axis of the dusty reflection nebulosity.  

The open question is what the nature of the wind is, is the ionized material
located in a bi-polar flow type of geometry or in an edge-on disk? Pending
velocity resolved high resolution imaging of the recombination lines, we
restrict ourselves to the following. The only velocity information known about
the large shell around the object is ambiguous. On the one hand Nedoluha \&
Bowers (1992\nocite{1992ApJ...392..249N}) find from their high resolution
imaging in the OH masers that the Northern part is tilted toward the observer
and the Southern part moves away. This picture would favour a bi-polar geometry,
as the velocity difference between the northern and southern components is too
large to be realistically explained by Keplerian motions. On the other hand,
Castro-Carrizo et al. (2007) find in their sub-arcsecond imaging of the CO
rotational line emission a largely symmetric appearance, but, intriguingly,
present a southern component approaching the observer, opposite to Nedoluha \&
Bowers data. Clearly, the nature of even the large scale structure of the
circumstellar data is not settled yet.

\section{Conclusions}
\label{concl}
We have presented VLTI/AMBER observations in the $K$-band of the post-Red
Supergiant transition object \IR, with the aim to spatially resolve the
Br$\gamma$ emission region. The observations were done
with three UT telescopes and consist of intermediate spectral
resolution spectra of the coherent and incoherent flux of \IR within
the AMBER field of view of 66\,mas and with an angular resolution of
6.5\,mas. The telescope configuration probed a range of 20\degr~in
position angle. The conclusions of this study are as follows.

\begin{itemize}

\item The AMBER visibility spectra show a decreased \Brg line
visibility for all three baselines.

It indicates that the line emitting region has been resolved. The continuum
visibilities on the other hand are not constrained
by this dataset. We fit the \Brg line visibilities with simple
intensity distributions, and find that a Gaussian distribution with a
FWHM of 3.3\,mas fits the data best. The derived diameter corresponds
to a distance of approximately $\rm 5\,R_{*}$ from the central
star.

\item The AMBER differential phase spectra show a change in phase for the U3-U1
  baseline at the \Brg transition.

This could indicate an offset of the photocentre projected on this baseline. New
measurements of better quality would help to characterize the nature of this
signal.

\item The AMBER flux spectra show the \Brg line of \IR in emission. 

It is stronger than reported by two previous studies. We have argued that this is due to the fact that most of the $K$-band excess
emission is located outside the AMBER field of view. This result is
consistent with the speckle-interferometric results presented
previously by Bl\"ocker et al. (1999), where they find the inner most
dust sphere to have a diameter of 70\,mas.
 
\end{itemize}
We have speculated on the geometry of the \Brg emitting region,
concluding that it would be elongated along the interferometer baseline if the
emission is optically thick. Optically thick emission seems however to be consistent with observed
line ratios. Additional measurements perpendicular to the baselines
presented here will help in further constraining the processes that shape the
environment of this massive star in transition to its final fate.

\begin{acknowledgements}
We would like to thank Ben Davies for useful discussions, and Andrew Clarke for
communicating his results on the near-infrared spectrum. This manuscript
benefited from insightful comments by an anonymous referee.
\end{acknowledgements}

\bibliographystyle{aa}


\end{document}